\documentclass[journal,twocolumn]{IEEEtran}
\usepackage{cite}
\usepackage[pdftex]{graphicx}
\usepackage{graphicx}
\usepackage{amsmath}
\usepackage{amssymb}
\usepackage{algorithmic}
\usepackage{array}
\usepackage{fixltx2e}

\usepackage{url}
\usepackage{epstopdf}
\usepackage[ruled]{algorithm2e}
\usepackage{multirow}
\usepackage{color}
\usepackage{booktabs}
\usepackage{subfigure}
\usepackage{epsfig}
\begin{document}

\title{ML-Assisted UE Positioning: Performance Analysis and 5G Architecture Enhancements}

\author{M. Majid~Butt,~\IEEEmembership{Senior~Member,~IEEE},~Anna Pantelidou~and~István Z. Kovács\\
%e-mail:majid.butt@nokia-bell-labs.com
%e-mail:{\{majid.butt,~istvan.kovacs,~anna.pantelidou,~teemu.veijalainen,~anil.rao,~mikko.saily\}@nokia-bell-labs.com}\\
%afef.feki@nokia.com, mll@es.aau.dk
 	
\thanks{This work has been presented in part at VTC-Spring 2020 \cite{Majid_VTC2020}.}	
\thanks{M. Majid~Butt,~István Z. Kovács,~Anna Pantelidou are with Nokia Bell Labs.
e-mail: \{majid.butt,~istvan.kovacs,~anna.pantelidou\}@nokia-bell-labs.com.
}
}

\maketitle

\begin{abstract}
Artificial intelligence and data-driven networks will be integral part of 6G systems. In this article, we comprehensively discuss implementation challenges and need for architectural changes in 5G radio access networks for integrating machine learning (ML) solutions.
As an example use case, we investigate user equipment (UE) positioning assisted by deep learning (DL) in 5G and beyond networks. As compared to state of the art positioning algorithms used in today's networks, radio signal fingerprinting and machine learning (ML) assisted positioning requires smaller additional feedback overhead; and the positioning estimates are made directly inside the radio access network (RAN), thereby assisting in radio resource management. In this regard, we study ML-assisted positioning methods and evaluate their performance using system level simulations for an outdoor scenario. The study is based on the use of raytracing tool, a 3GPP 5G NR compliant system level simulator and DL framework to estimate positioning accuracy of the UE. We evaluate and compare performance of various DL models and show mean positioning error in the range of 1-1.5m for a 2-hidden layer DL architecture with appropriate feature-modeling. Building on our performance analysis, we discuss pros and cons of various architectures to implement ML solutions for future networks and draw conclusions on the most suitable architecture.
\end{abstract}


\IEEEpeerreviewmaketitle

%%%%%%%%%%%%%%%%%%%%%%%%%%%%%%%%%%%%%%%%%%%%%%%%%%%%%%%%%%%%%
\section{Introduction}
\label{sec:intro}
\subsection{Intelligence in Beyond 5G Networks}
A lot of attention has been drawn in the last few years around the 5th generation (5G) New Radio (NR) cellular systems, which has been under standardization by the 3rd Generation Partnership Project (3GPP). 5G promises ubiquitous connectivity of a massive number of heterogeneous services.
Users in a 5G network may have very different performance requirements in terms of data rates, reliability and latency. To support those, 5G networks are designed to be very flexible and sophisticated, but also highly complex. This renders optimization of different network functions difficult. Artificial intelligence (AI) and Machine Learning (ML) in particular have been recently investigated as promising techniques that can be used to optimize network functions by introducing an additional data-driven intelligence to the network \cite{Khalid:2019}.

Network virtualization and network softwarization solutions in 4G networks can support data-driven intelligent and automated networks to some extent.
In 5G and beyond systems, network intelligence is envisioned to be end-to-end. The user equipment (UE) devices need to be smarter, environment and context aware, and capable of running ML algorithms. The radio access network algorithms and radio resource management (RRM) functions can exploit network intelligence to fine tune network parameters to reach close-to-optimal performance.
More dynamic network topology and more flexible self-organization need to be integrated in the network to support highly versatile network traffic. The era of network virtualization and network softwarization in 4G and 5G leads to the era of smart, intelligent and automated networks of the future in 6G.

\begin{figure*}
\centering
  \subfigure[Conventional Positioning]
  	{\includegraphics[width=2.5in]{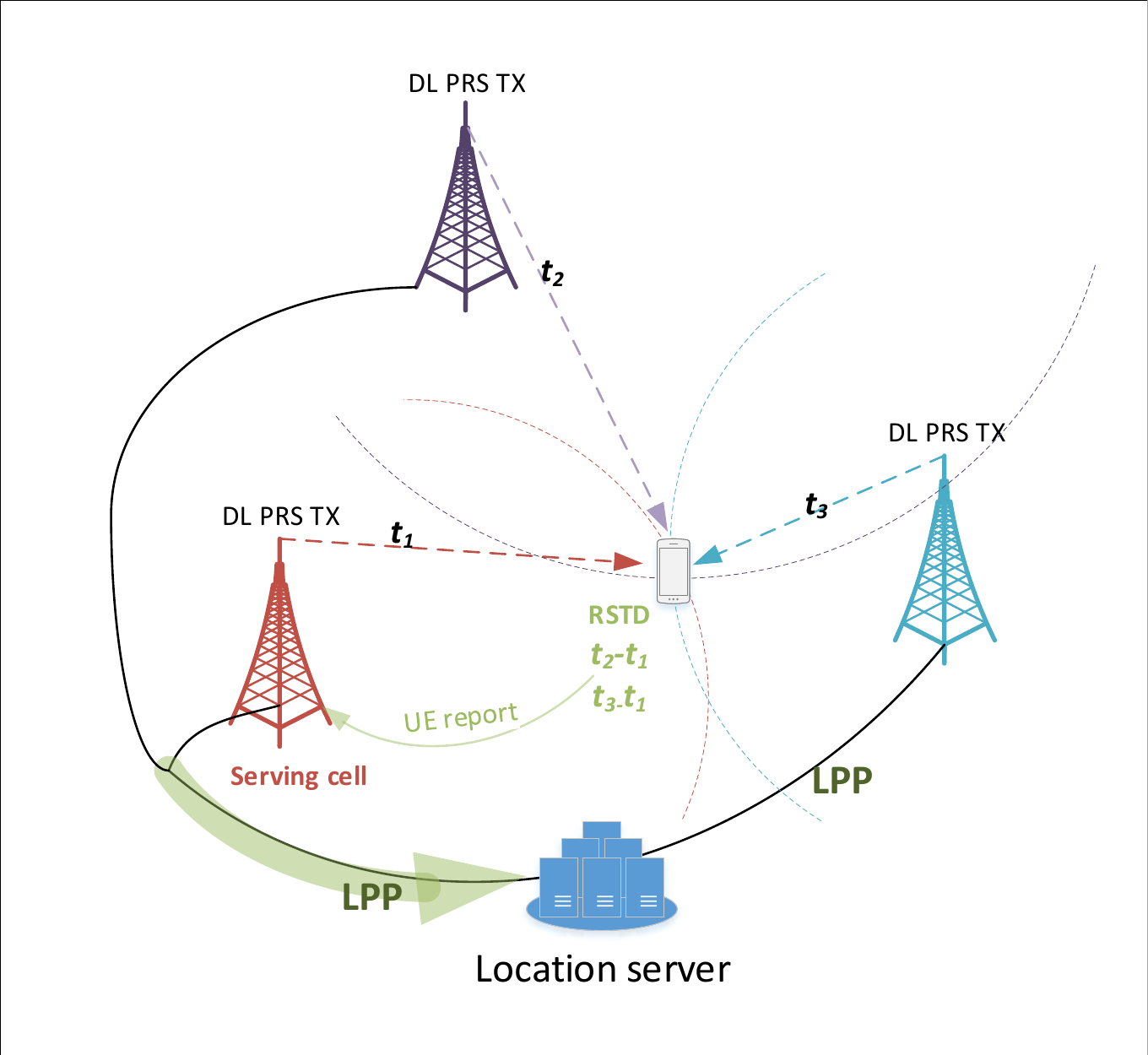}
  \label{fig:CM}}
 \subfigure[ML-assisted Positioning]
  	{\includegraphics[width=2.5in]{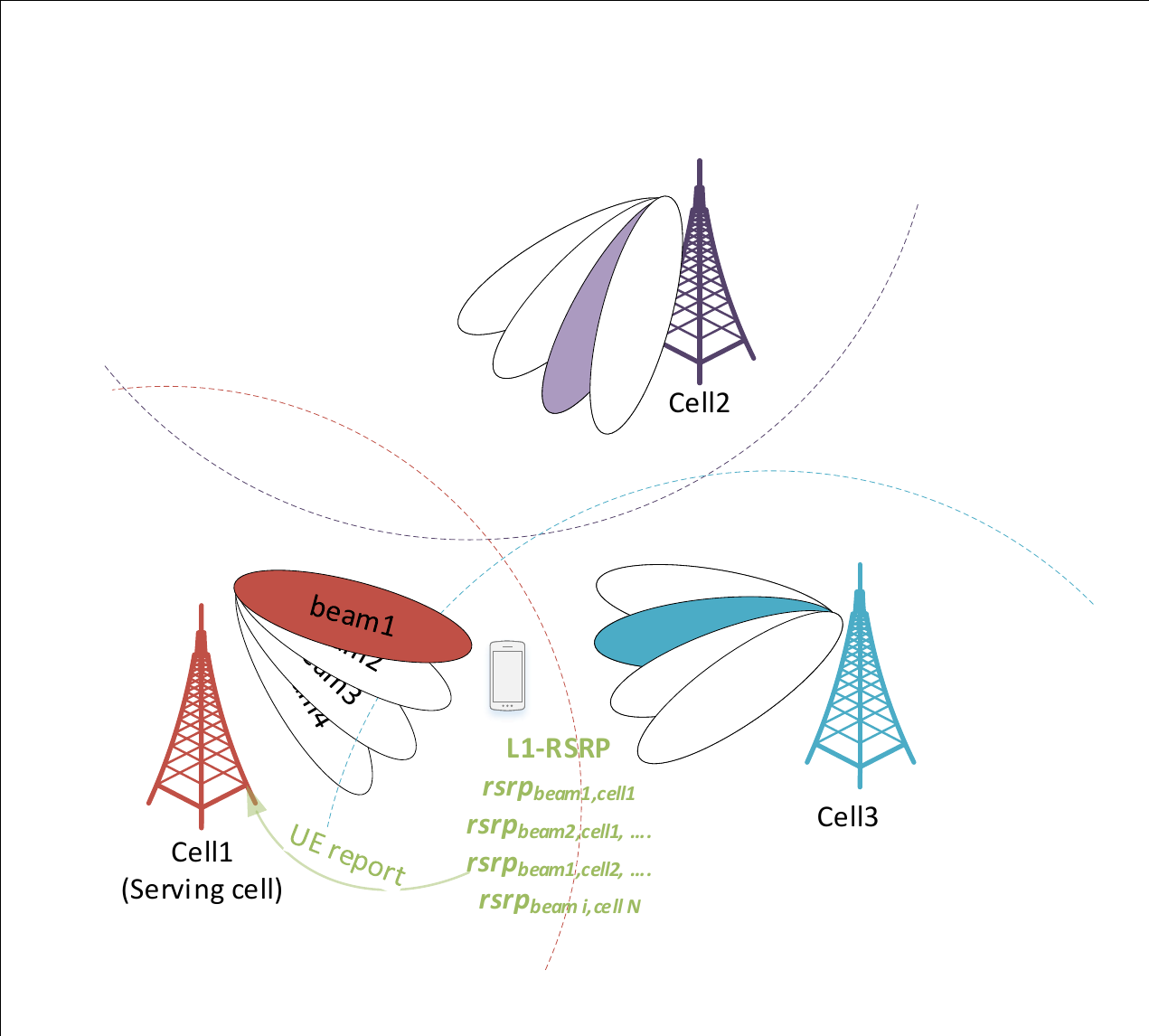}
  \label{fig:ML}}
   \caption{In conventional positioning solutions, beam RSRP values as well as angle of arrival is reported back by a UE to gNB. Our proposed ML solution exploits just RSRP values reported by the UE.}
	\label{fig:positioning_methods}
\end{figure*}

\subsection{UE Positioning: Motivation and State of the Art}
5G technologies introduce a new paradigm to connectivity and mobility by using accurate location information. Positioning and localization studies have been conducted for various applications and devices for many decades such as global positioning systems, industrial internet of things, autonomous driving, indoor maps and so on \cite{positioning_NR,positioning_LTE}. Positioning applications are found from small indoor area to wide landscapes of satellite coverage. As mobile handsets are widespread, users start utilizing localization services through cellular networks. In the 5G era, even higher number of device types appear such as sensor devices, robots and vehicles; all operating based on highly accurate location information.

Positioning accuracy is one of the most important factors to evaluate localization system performances. Federal Communications Commission (FCC) in the United States has set the E911 regulatory performance requirements \cite{FCCE911} as well as further accuracy requirements defined tightly by commercial systems \cite{positioning_NR}. FCC E911 requires horizontal error less than 50m and vertical error less than 5m for $80\%$ of UEs \cite{FCCE911}. In commercial systems like 3GPP, requirements are tighter than E911 ones. To reach these levels, novel enhanced positioning methods need to be investigated.

In order to achieve such goals of high accuracy and measurement efficiency, various positioning methods have been deeply studied. In localization systems, usually a positioning method is defined based on a particular type of measurement metric. As illustrated in Fig. \ref{fig:CM}, timing-based techniques take advantage of knowledge of time-differences due to the speed of RF signals for distance calculation between a transmitter and a receiver. The measurement matrix is estimation of the time of arrival (TOA) of signals. Angle-based techniques take advantage of transmit beamforming of signals and differences of phase across receive antenna elements to determine the angle between a transmitter and a receiver. There is a lot of literature available on UE positioning in 5G based on UE signal time and angle measurements or a combination of them, e.g., \cite{Ennasr,Talvitie_2019,Liu:Network,Wymeersch_2017}. A detailed overview of positioning technology and road map to 5G positioning architecture is provided in \cite{Liu:Network} while Wymeersch et al discuss positioning for vehicular networks in 5G era \cite{Wymeersch_2017}.

As the NR physical layer design inherently supports beam-based Reference Signal Received Power (RSRP) measurement reports from the UE for multiple beams, we study how well ML-based approaches work to determine the UE position directly from these measurements as illustrated in Fig. \ref{fig:ML}. During the ML inference, the only inputs needed are measurements the UE is providing to its serving cell, hence the positioning estimation can be done using an ML inference engine in the baseband of the UE's serving cell, avoiding the need for the location server framework which resides outside the radio access network and therefore, efficiently supporting a single-cell based positioning method.

ML techniques have been used to solve various problems in communication systems. References \cite{Simeone:TCN,Shea:TCN_2017,DL_survey:2019} survey some exciting works on use of ML algorithms in the field of wireless communication and networking. Some recent works apply ML techniques for localization problems. In \cite{Xiao} autoencoder-based indoor localization method is studied that provides high-performance 3-D positioning in large indoor places for bluetooth low energy networks. The authors in \cite{Prasad_TWC:2018} propose a machine learning approach based on Gaussian process regression to position users in a distributed massive multiple-input multiple-output (MIMO) system with the uplink received signal strength (RSS) data.
The simulation results show that the proposed method improves performance in massive MIMO scenarios. In \cite{Ye_IEEEAceess_2017}, neural networks are used on input of fingerprint data obtained from channel characteristics and geographical locations and median error of 6 and 75 meters is reported for indoor and outdoor scenarios, respectively. Similar approaches for UE positioning using neural networks (NNs) are investigated in \cite{Wu_electronics:2019, Q_zhang_Eurasip2020, Lee_ICASSP:2018, Magnus, Gante_2020} as well.

Most of these works discuss ML-based positioning in different settings than ours. For instance, work in \cite{Xiao} is limited to bluetooth networks, \cite{Wu_electronics:2019, Q_zhang_Eurasip2020} focus mainly on data from Wifi networks, while results in \cite{Lee_ICASSP:2018} are limited to the case where data is available reliably from three base stations, none of them is a limitation of our work. The authors in \cite{Ye_IEEEAceess_2017} use more practical settings for their evaluation but their feature selection is different and evaluation is limited to indoor scenarios unlike this work, where evaluation is carried our for more challenging out door scenarios by making use of raytracing. The works in \cite{Magnus} and \cite{Gante_2020} evaluate UE positioning in the context of 5G networks but evaluations are very limited in terms of details on feature selection and no insight is provided on challenges associated with deployment of ML based positioning in 5G networks. The authors in  \cite{Prasad_TWC:2018} provide fundamental results on error bounds using ML and no evaluation results are provided for the practical cellular networks. Furthermore, we dedicate second part of the paper to network architecture aspects, including measurement collection, ML module hosting and associated challenges in practical deployment of the solution in 5G networks. This study is integral part of 5G networks going forward and missing in literature.

\subsection{Contributions}
Recent literature on the usage of ML in wireless networks emphasizes on fundamental communication problems \cite{Zappone2019WirelessND}, potential use cases \cite{Koivisto_2017} and associated challenges; but there is a lack of studies on the larger scale performance analysis at system level and the challenges associated with ML algorithms implementation in 5G networks. Standardization bodies, such as International Telecommunication Union \cite{ITU_recommend3172}, and 3GPP (Rel 17 and Rel 18) have started working on identifying requirements for the future networks.

This article complements the standardization efforts and aims at filling this gap by studying performance of UE positioning as an example use case, and discussing architecture level aspects for implementing ML-assisted solutions in 5G and beyond networks.

Our contributions are summarized as follows:
\begin{itemize}
  \item We provide a comprehensive system level study starting from generating raytracing data, using system level simulator to generate RSRP fingerprinting for geographical $(X,Y)$ coordinates; and finally applying ML techniques to train and validate UE positioning accuracy.
  \item We study various input features that contribute to improve positioning accuracy and conclude that a combination of RSRP values from the serving and neighboring cell beams provides the best results. Then, we evaluate the performance of network level and cell-specific training topologies and quantitatively show that cell-specific training outperforms network level training. For the sake of comparison, we evaluate decision tree regression ML technique to evaluate the performance.
  \item We identify main requirements to implement ML solution for UE positioning use case. In particular, we identify the data source for the ML model host and analyze how data could be gathered in 5G-RAN architecture to train the ML model and subsequently use it for inference. Then, we identify ML model hosts in relation to 5G-RAN architecture and describe pros and cons for the proposed solutions with respect to data availability and overhead cost.
\end{itemize}

The rest of this paper is organized as follows: after describing 3GPP system activities in relation to UE positioning in Section \ref{sec:3GPP}, we proceed with details of set up for our case study in Section \ref{section:positioning}. Performance of the ML approach is evaluated and discussed in Section \ref{sec:perfromance}. Section \ref{sec:arch_5G} and \ref{sec:arch_support} discuss 5G architecture level support to run ML solution for this use case and associated design challenges. Section \ref{sec:conclusions} concludes with the main contributions of the paper.

\section{3GPP Standardization on Positioning}
\label{sec:3GPP}
3GPP positioning feature has been supported in Rel-9 LTE networks; several RF based positioning methods have been introduced, and advanced techniques are under investigation for NR in Rel-16 and Rel-17. Recent 3GPP NR positioning study conclusions have been stated in \cite{positioning_3GPP}. There has been discussion of many different possible positioning solutions in the study item, conclusions have been made that the 3GPP NR systems should support solutions of observed time difference of arrival (OTDOA), uplink time difference of arrival (UTDOA), angle information, Multi-Cell Round Trip Time (Multi-RTT), and enhanced cell-ID (E-CID) in NR Rel-16 \cite{qualcom2014}.
For the proposed solutions, 3GPP is under discussion to define network measurements through DL or UL radio link.
%
%OTDOA requires UEs to measure DL-TDOA. UEs measure TDOA based on DL positioning reference signals from the neighbor and serving cells. A location server collects TDOA measurements and compute a UE location. UTDOA requires gNBs to measure ToA based on UL positioning reference signal.
%Also, angle measurement is an important information for positioning.

One of the major changes in NR compared with legacy cellular networks is the addition of the millimeter wave (mmWave) frequency band.
NR also provides support for beam-based operations. During the beam search operation, the angle can be obtained easily in terms of downlink angle of departure (DL-AoD) and uplink angle of arrival (UL-AoA). While UEs and gNBs search for the best beam for radio link, DL and UL RSRP is measured to evaluate signal quality of a beam.
Received power measurement has been considered for positioning fingerprint in wireless systems \cite{pattern_matching_3GPP}.

%In LTE system, E-CID method is widely used for positioning based on the received signaling power. The location server constructs a fingerprint for measurement-to-location mapping, but it needs to involve measurements from many cells for high accuracy.

\begin{figure*}
\centering
  	\includegraphics[width=4.5in]{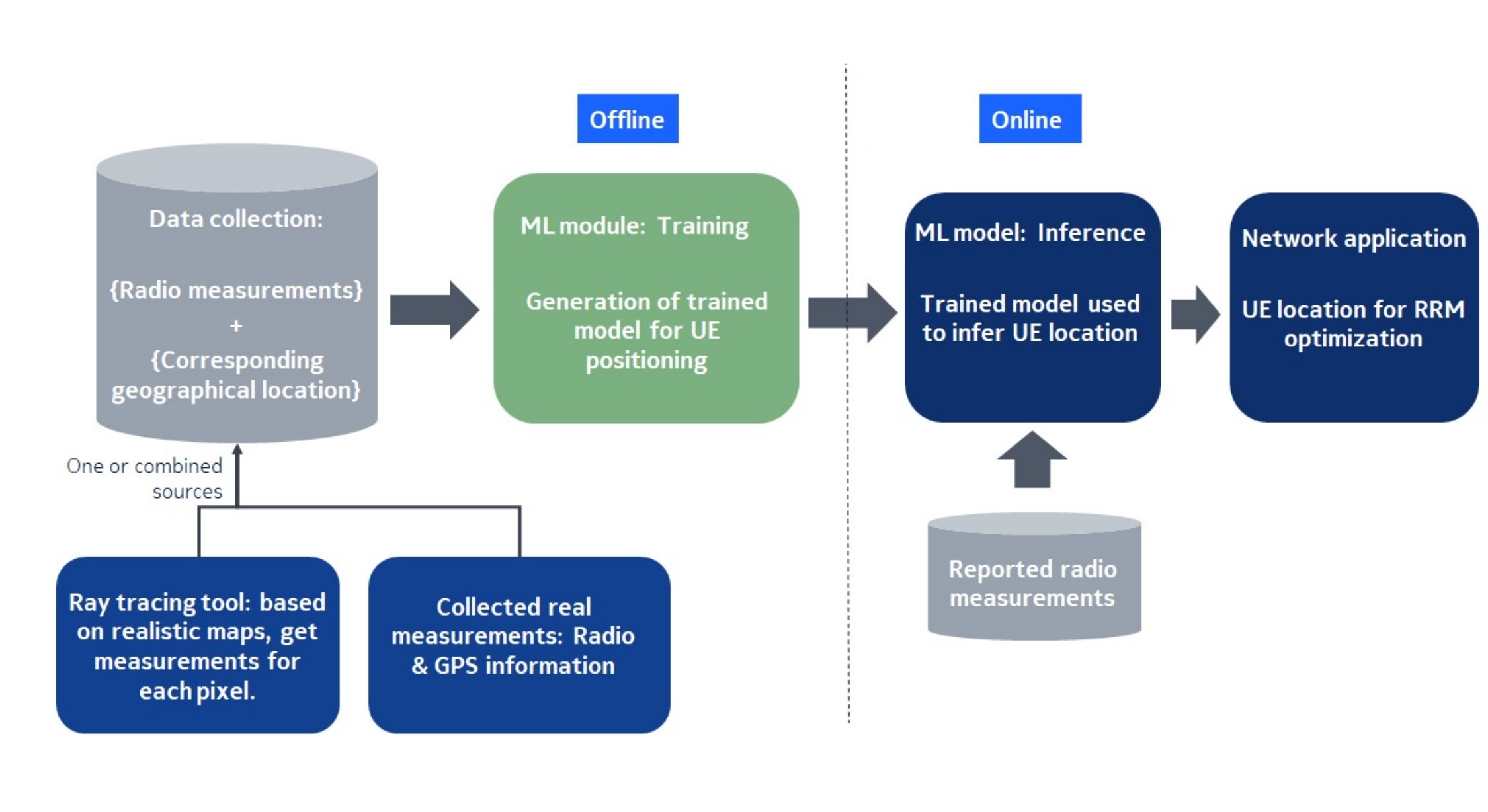}
   \caption{Schematic diagram for our proposed ML-assisted UE positioning framework. Offline data pre-processing and ML training is performed while online part performs inference.}
	\label{fig:model}
\end{figure*}

%As the NR system is required to construct measurement per beam, we investigate NR received signal power based positioning methods with the concept. We evaluate ML approaches for computing positioning accuracy for the UE to meet positioning accuracy requirements. Based on RSRP of the beams (without angle/time of arrival calculation), we determine positioning accuracy of the UE.

5G technologies introduce a new paradigm to connectivity, location awareness and RRM by providing accurate UE location information to network functions. In future 3GPP releases, positioning enhancements are envisioned to achieve different levels of positioning accuracy for different applications. Positioning requirements in 3GPP are under discussion with horizontal positioning accuracy for general commercial use cases (80\% of users) with less than 1m accuracy and the positioning latency below 100ms \cite{TS22.261}. For industrial IoT use cases, the position accuracy of less than 20cm is required with a desired latency in the order of 10ms \cite{TS22.804}.

3GPP Release-17 and beyond is further  evolving to improve accuracy and latency with advanced technologies in specific scenarios. 3GPP has interests in centi-meter accuracy positioning measurements through carrier phase measurement or wideband positioning measurements \cite{positioning_NR,positioning_NR_Rel17}. Positioning using machine learning is one of key topics as well. Positioning service latency is another important factor to evaluation positioning system especially for vehicle and industrial use cases. As enabling technologies, V2X and sidelink supports ad-hoc positioning measurements, and on-demand positioning reference signal (PRS) supports rapid scheduling of PRS and measurements effectively.

Achieving better accuracy requires more radio resources and signal bandwidth, computational power, more overhead in collecting more measurements and more delay in processing this information. Thus, positioning can be treated as a service with different quality of service (QoS) attributes. A request for UE location may be conveyed with a required QoS, which includes the desired location accuracy (vertical and/or horizontal) as well as the response time. In fact, the location service (LCS) can allow the LCS client to specify its required quality of service parameter. In 3GPP standard, two different LCS QoS class options are defined \cite{3GPP_Afef}. In \emph{assured class}, the requirements for estimated UE position are strict in terms of accuracy and response time. If the obtained result does not fulfill the indicated requirements, then the response will be discarded by the LCS service. In the \emph{best effort class}, the requested location QoS is not strict and the LCS service should receive the information, even if it does not fulfill the indicated requirement. The selection of the most suitable positioning method, e.g., AoA, ToA based methods; and its related parameters should account for this QoS information when available.

%\begin{itemize}
%  \item \textbf{Assured:} The requirements for estimated UE position are strict in terms of accuracy and response time. In this case, if the obtained result does not fulfill the indicated requirements, then the response will be discarded by the LCS service.
%  \item \textbf{Best effort:} The requested location QoS is not strict. The UE position can be estimated with the available method and the LCS service should receive the information, even if it does not fulfill the indicated requirement.
%\end{itemize}

%Typically, when an ML method is applied, the QoS class can impact the selection of the most appropriate inputs for the ML model such as the number of neighboring cells measurements and the measurement type (e.g. received power level, delay, AoA).
%Besides, this QoS information can be used to evaluate the results of a previous request to validate accuracy of the trained model and decide if model retraining is required.

\section{Positioning in 5G Systems: A Case Study}
\label{section:positioning}

\subsection{System Setup}
\label{sec:system_setup}
We first describe the system set up for our study. As shown in Fig. \ref{fig:model}, the system set up is divided into the following main components:

\begin{enumerate}
  \item {\textbf{Raytracing}: A raytracing software is used to generate raytracing data as a function of $(X,Y)$ coordinates, and full 3-D multipath ray data is generated for each point. In this case, an outdoor urban database and the Intelligent Ray Tracing algorithm were configured.}
  \item {\textbf{System Level Simulation}: As the NR system is required to construct measurement per beam, we investigate NR received signal power based positioning method. Based on the raytracing data generated for channel modelling, system level simulation of the 3GPP 5G NR air interface including beamforming is run to produce beam RSRP values for each UE. The details of the parameters used in system level simulations are provided in Table \ref{tab:winprop} and Table \ref{tab:system_level}. The serving as well neighboring beam RSRP values serve as fingerprinting for the under-study area based on raytracing data.}
  \item{\textbf{Offline ML Training}: Based on the data received for the system level simulator, offline ML training is performed. We select beam RSRP values as input features and train ML model with feed-forward neural network as well as decision tree algorithms.}
  \item{\textbf{Online ML Inference}: Finally, trained ML model is numerically evaluated to determine UE location. \textit{Euclidean distance} between the estimated and actual UE position, which is assumed to be known, serves as the accuracy measure for the estimated UE position.}
\end{enumerate}
The last block in Fig. \ref{fig:model} points toward possible use cases for application of accurately estimated positioning data.

\begin{figure}
\centering
  	\includegraphics[width=2.7in]{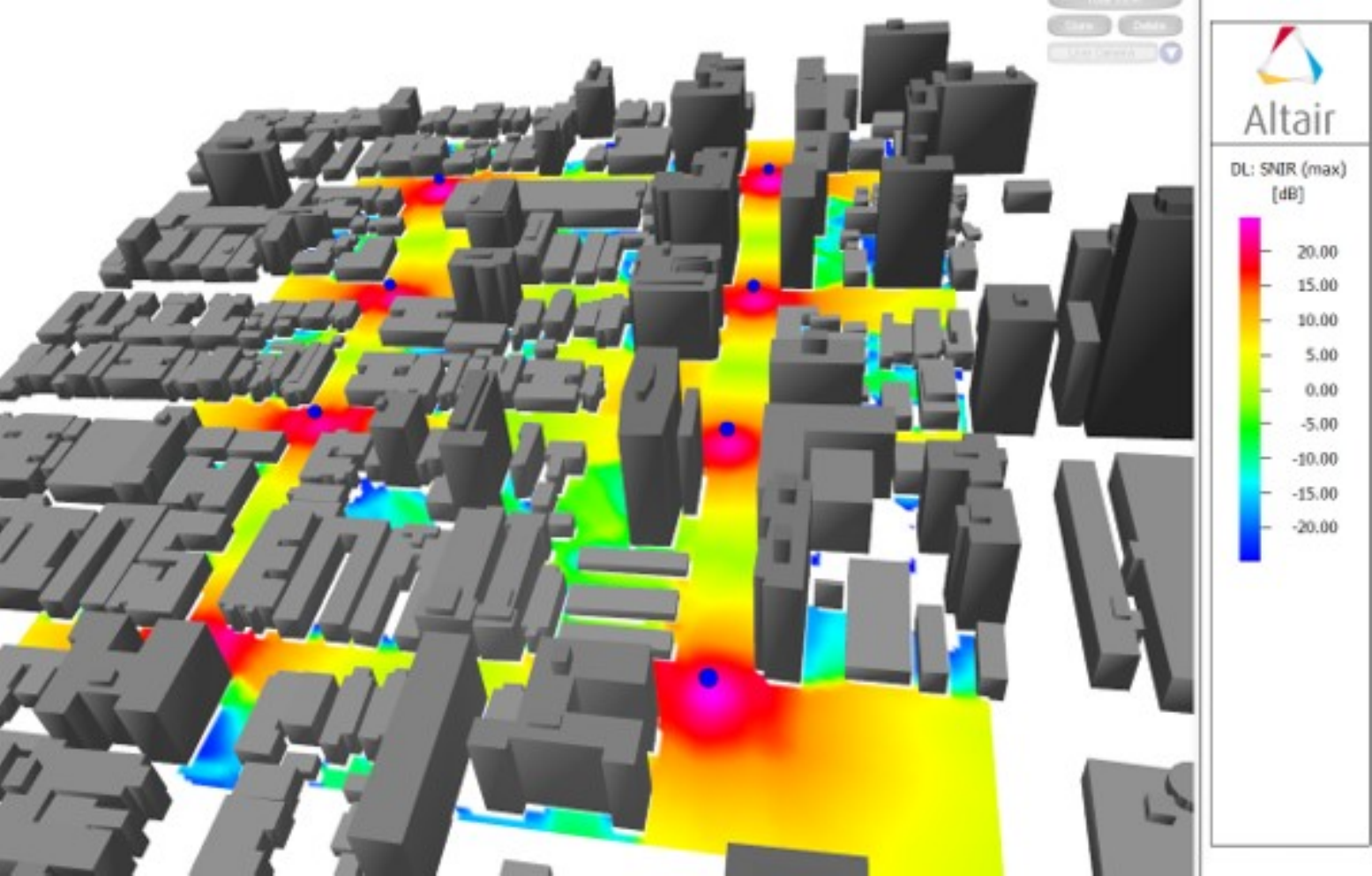}
  %\vspace{-0.2cm}
   \caption{8 sites are placed at 8 corners of Lincoln Park, Chicago for the purpose of raytracing data generation.}
	\label{fig:Raytracing}
%\vspace{-0.5cm}
\end{figure}

Raytracing is performed for a section of Lincoln Park, Chicago. Eight sites are placed at each street corner as shown in Fig. \ref{fig:Raytracing}. In this study, we have confined ourselves to study of positioning accuracy for the UEs that have a Line of sight (LoS) connection to their serving cell. We observed that positioning accuracy using this ML-assisted method in the mmWave propagation environment is severely affected when LoS is not available to the UE. However, this may not be a big issue in future networks where very dense mmWave deployments are planned and LoS non-availability chances will not be that high. In our study, $62\%$ of the UEs have LoS available and we consider only those users for ML training and testing. For ML training purpose, we only consider RSRP feature for the beams in current time slot, i.e. a snapshot based approach is employed. The methodology used in this study provides an efficient option to evaluate feasibility of this approach compared to drive testing.

We use two parameters, mean and standard deviation denoted by $\bar{E}$ and $\sigma$ respectively, of the Euclidean distance between the estimated training/test UE position and the actual UE position and define them as,
\begin{eqnarray}
  \bar{E} &=& \frac{\sum_{n=1}^N x_n}{N} \\
  \sigma &=& \sqrt{\frac{1}{N}{\sum_{n=1}^N (x_n-\bar{E})^2}}
\end{eqnarray}
where $x_n$ is the value for UE Euclidean distance for $n^{th}$ sample.

\subsection{Summary of Deep Learning Architecture}
Deep learning is one of the powerful ML techniques to solve classification/regression problems. Deep feedforward networks are advanced neural networks (NNs) with one or more hidden layers of neurons between input and output layer.
%
%In \emph{feedforward NNs}, information flow from input $x$ to output $y$ through intermediate layers and there is no feedback, contrast to Recurrent NNs where information is fed back as well.

%\begin{figure}
%\centering
%  	\includegraphics[width=3.3in]{./pics/NN}
%   \caption{Schematic diagram for a DNN. Input and output layer width/size is defined by the input and output requirements while depth and width of DNN are hyper parameters need to be optimized.}
%	\label{fig:NN}
%\vspace{-0.4cm}
%\end{figure}

A feedforward deep NN (DNN) has an input layer, which is connected with output layer through hidden layers. The hidden layers represent \emph{depth} of the network, and 'deep learning' term arises from this. The number of neurons in each layer represent \emph{width} of an NN where neuron is a computational unit that represents a vector-to-scalar mapping. An activation function, such as 'tanh' and 'Rectified Linear Units (ReLU)', is used to compute element (neuron) wise non-linear transformation of learned parameters 'weights' and 'bias', which is transferred to next layer. Goal of a DNN is to learn the function that maps input $x$ to output $y$. Input vector $x$ is termed as \emph{feature vector} in DL literature. Architecture of a neural network defines how many layers a NN has, how many neurons per layer are required and how layers are connected. This is a stochastic process and termed as 'hyperparameter' optimization.\footnote{Hyperparameter optimization is one of the main tasks in ML and include optimization of several parameters by experimentation.}

%A deep NN is first \emph{trained} on large number of examples/measurements provided where solution $y$ is provided for the input $x$, called supervised learning.
For supervised learning, the term 'Loss' is defined as the difference between the computed output and actual output parameterized by some function, such as minimum squared error (MSE), entropy function, etc.
Based on the training data, DNN tries to learn mapping function that minimizes 'loss' as training progresses.
%Once the function is computed, it is 'tested' against the data not part of training to validate the computed DNN model.
During inference, test error is computed by an appropriate metric to compute accuracy of the trained DNN.

In ML, 'overfitting' and 'underfitting' are two important concepts. 'Overfitting' represents the situation when DNN is over-trained and it becomes hard for the computed model to be generalized for the unknown testing examples. To overcome this problem, 'early stopping' and 'regularization' is used. On the other side, 'underfitting' represents the situation when both training and testing error are large \cite{Goodfellow-et-al-2016}.
%This shows that DNN model has not been trained well and the computed function cannot be generalized accurately to the unknown examples. This can happen because of lack of training data or not training the DNN for sufficiently enough time.
Analysis of training and testing error provides good indication of accuracy of DNN model and helps designer take right actions to improve the model. Collecting more data and/or hyperparameter optimization are the well-known solutions.

%%%%%%%%%%%%%%%%%%%%%%%%%%%%%%%%%%%%%%%%%%%%%%%%%%%%%%%%%%%%%%%%%%%%%%%%%%%%%%%%%%%%%%%%%%%%%%%%

%Machine learning assisted positioning methods can contribute to enhance positioning accuracy and provide various levels of accuracy and latency for different applications.
Unlike conventional positioning methods based on ToA and AoA information, ML assisted positioning does not necessarily require measurements from the neighbouring gNBs for triangulation. ML-assisted algorithms can be used to compute UE position based on RF-fingerprinting of UE location and the RSRP of the beams serving the UE.
In the next section, we present results based on a case study conducted in a millimeter wave (mmWave) outdoor scenario.

\section{Performance Evaluation}
\label{sec:perfromance}

\begin{table}
%\vspace{-0.2cm}
\begin{center}
\footnotesize
\caption{Raytracing Parameters}
%\vspace{-0.3cm}
\begin{tabular}{ll}
\toprule
Parameter & Value\\
\midrule
Number of sites							& 8\\
Access Point height & 10m\\
Inter-site distance& 110m vertical, 200m horizontal\\
Carrier frequency & 28 GHz\\
Resolution of data points & 1m (123,768 locations sampled)\\
\bottomrule
\end{tabular}

\label{tab:winprop}
\end{center}
%\vspace{-0.4cm}
\end{table}

\begin{table}
\begin{center}
\footnotesize
\caption{System Level Simulation Parameters}
%\vspace{-0.3cm}
\begin{tabular}{ll}
\toprule
Parameter & Value\\
\midrule

Number of SSB beams &32\\
Number of gNBs 	 					& 8 having 3 sectors each\\
DL Tx Power per sector & 30 dBm\\
Element Azimuth Beamwidth&65 degrees\\
Element Elevation Beamwidth&65 degrees\\
Element gain & 8dBi\\
Front2Back & 30dB\\
Mechanical downtilt & 5 degrees\\
\bottomrule
\end{tabular}

\label{tab:system_level}
\end{center}
%\vspace{-0.4cm}
\end{table}

\begin{figure*}
\centering
  \subfigure[Example 1]
  	{\includegraphics[width=3.5in]{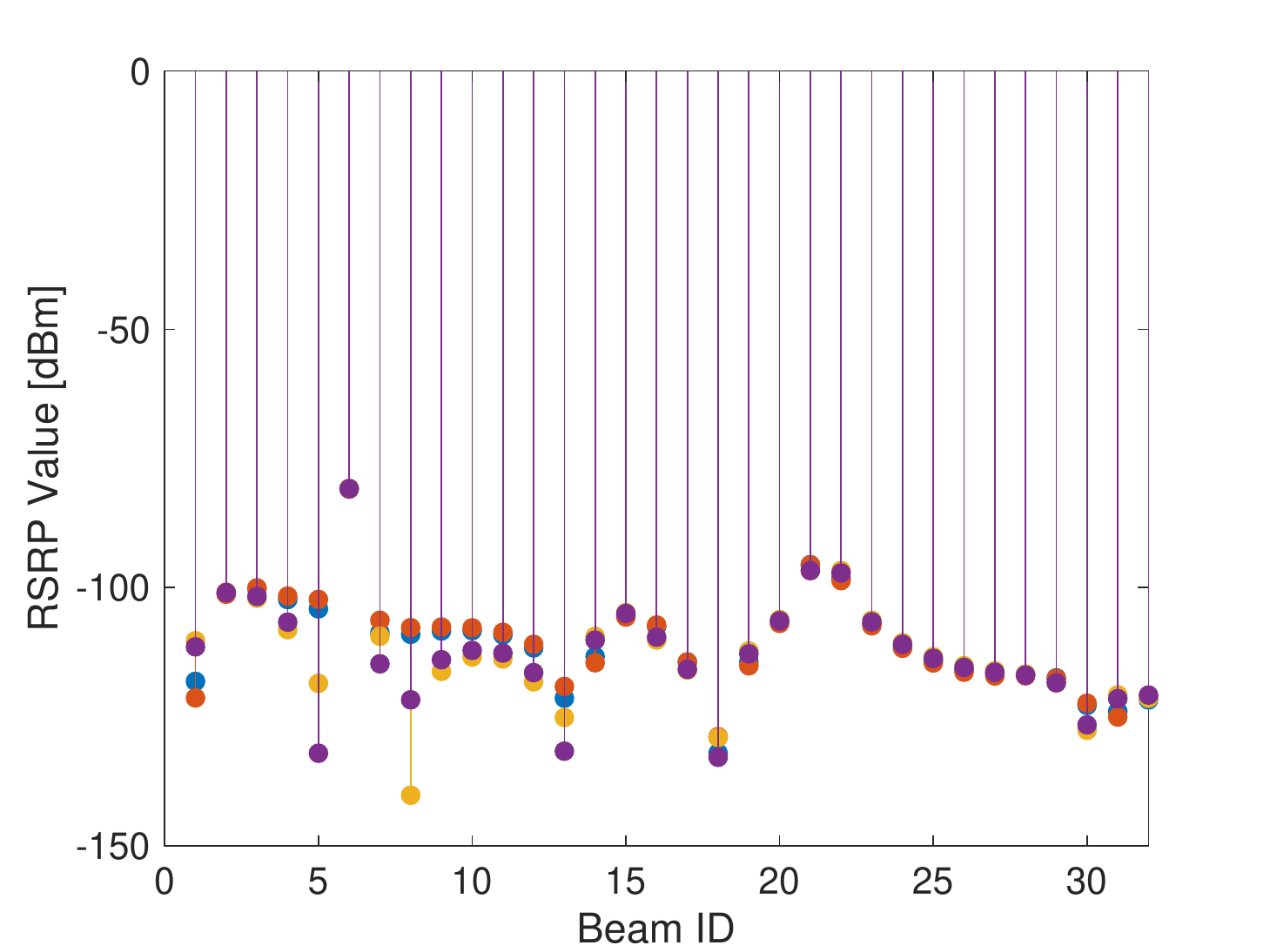}
  \label{fig:nonoverlape}}
 \subfigure[Example 2]
  	{\includegraphics[width=3.5in]{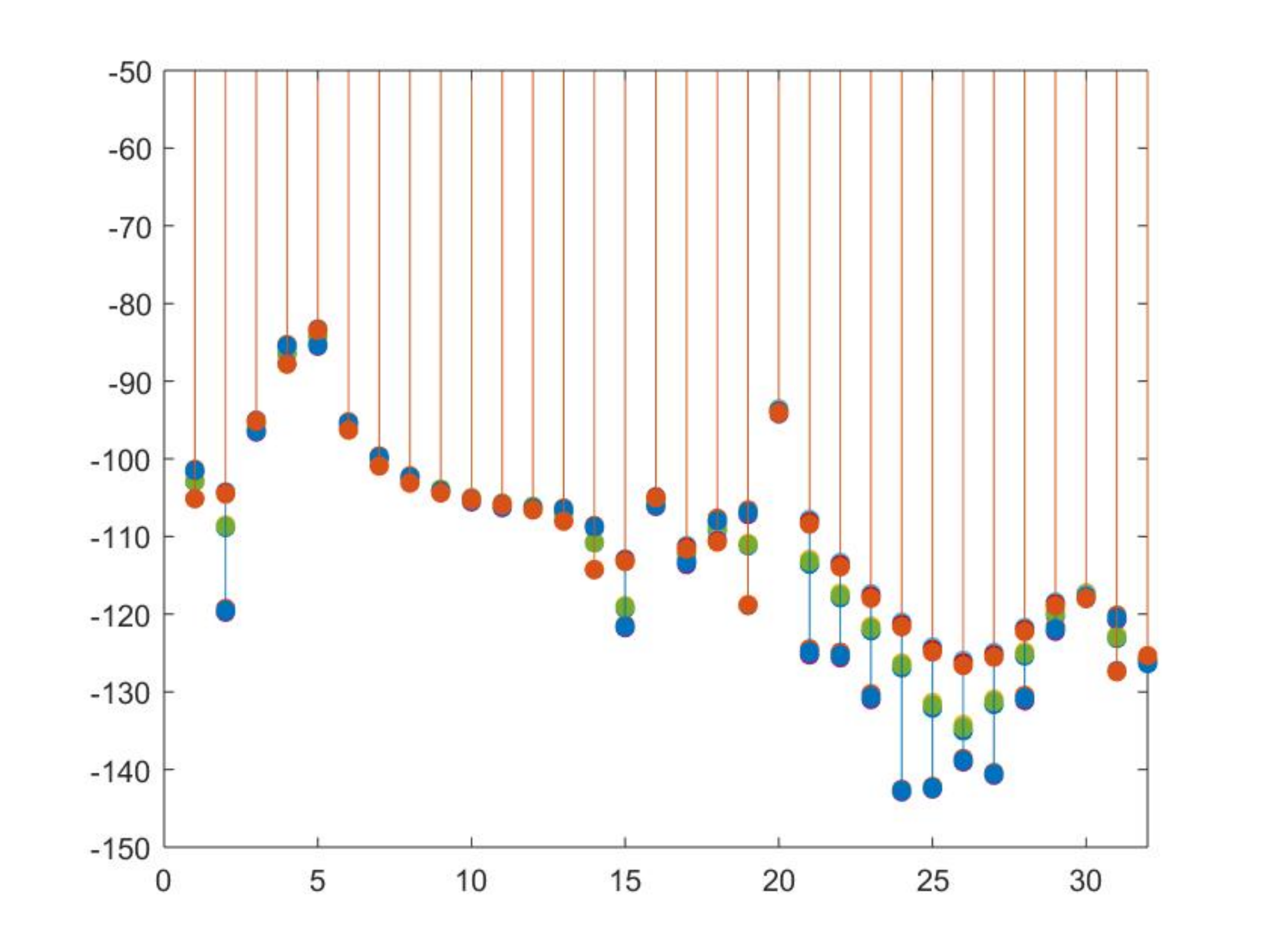}
  \label{fig:overlape}}
  %\vspace{-0.2cm}
   \caption{Pattern for RSRP values for the 32 beams received at the UEs in close proximity.}
	\label{fig:RSRP}
%\vspace{-0.4cm}
\end{figure*}
Parameters used for raytracing in raytracing software and system level simulations are summarized in Table \ref{tab:winprop} and \ref{tab:system_level}, respectively. In particular, Table \ref{tab:winprop} shows the parameters used for raytracing module with 8 sites, each with heigh 10m, as shown in Fig. \ref{fig:Raytracing}. Resolution of raytraced data points is 1m, i.e., raytracing area is sampled at interspacing of 1m.

In Table \ref{tab:system_level}, parameters used in system-level simulations are shown. The coverage is beam-based in 5G, not cell based. There is no cell-level reference channel to measure the coverage of the cell. Instead, each cell has one or multiple Synchronization Signal Block Beam (SSB) beams and the UE can select candidate beams after beam sweeping. The antenna panel used for system level simulations was a $16\times16$ cross-pol array, formed of 16 azimuthal beams and 3 elevation beams and grid-of-beam comprises 16 non-uniform beams to cover the full cell. Further details, on physical antenna parameters are provided in Table \ref{tab:system_level}.

For evaluation of DL module, the parametric settings are explained as follows. 90$\%$ of the labeled data is used for training and $10\%$ is set aside for validation. Other parameters include, Batch size = 32, 'tanh' activation function for input and hidden layers, linear activation function for output layer, 'adam' optimizer, loss metric = 'MSE'. We used Keras API with TensorFlow backend. 'Early Stopping' based on the 'loss' in the training data is set to prevent overfitting while maximum number of epochs is set to 500. Data is normalized using mean and standard deviation normalization.
%\footnote{Explanation details of these parameters can be found in common ML literature, e.g., \cite{Keras}.}.

\subsection{Decision-Tree Regression}
\begin{figure}
\centering
  	\includegraphics[width=3.0in]{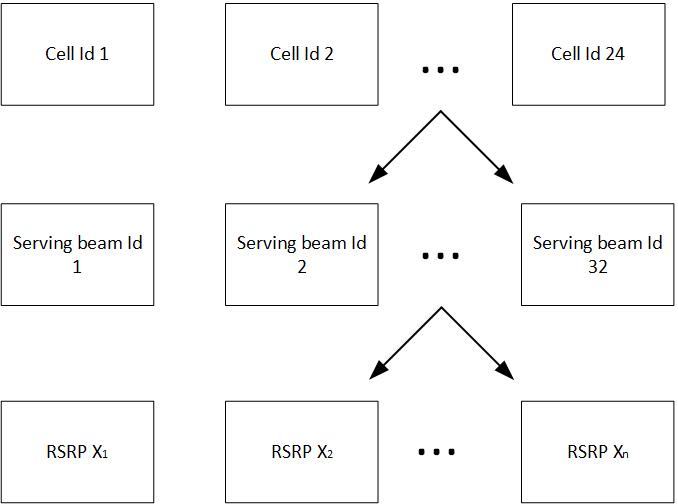}
   \caption{Block diagram for the tree like feature structure considered in this work.}
	\label{fig:tree}
\end{figure}

For the baseline ML solution, we use Decision Tree Regressor algorithm to evaluate performance. Decision Tree is a supervised learning method used for classification and regression for the problems where feature vectors have 'tree like' structure. The goal is to train a model that predicts the value of a target variable by learning decision rules inferred from the data features. Decision trees are simple to implement and results can be easily visualized and analyzed. Our feature vector (explained later in this section) has a 'tree like' structure as illustrated in Fig.~\ref{fig:tree} and suits well to regression problem at hand. The UEs are first subdivided based on cell IDs.
Further divisions are done based on beam IDs, and finally RSRP values help to differentiate between UEs with the same beam and cell ID. Later, in this section, we compare the Decision-Tree Regressor algorithm based performance evaluation with the DL based results.

\subsection{Network-Level DL Training}
\label{sec:network_level}
We first study the case where ML training is performed over all the data available from 24 sites at one positioning server. Current 3GPP standards support UE reporting for up to 4 strongest beams from the serving cell. In Table \ref{tab:comparison bb}, we show training and testing accuracy for the case when RSRP values for 4 strongest beams from the serving cell, corresponding beam IDs and cell ID is the input for the input layer of the DNN. This will serve as a baseline. For the 9 input features ($4\times$ beam ID, $4\times$ RSRPs, Cell ID), we train our DNN. We use only one hidden layer to get some idea on suitability of features and vary the number of neurons for each experiment\footnote{Multilayer NNs improve accuracy as we show later, however we confine ourselves to single hidden layer in the beginning to get more insight on the other parameters influencing our model.}. Estimation accuracy improves with increasing number of neurons and we get mean test accuracy of approximately 5m for very large number of neurons.

\begin{table}
\caption{Positioning Accuracy for 4 Strongest beams from the Serving Cell}
%\vspace{-0.4cm}
\label{tab:comparison bb}
\begin{center}
\begin{tabular}{|c|c|c|c|c|}
\hline
  No. &Training $\bar{E}$  &Training $\sigma$  &Test $\bar{E}$ &Test $\sigma$ \\
  Neurons& (m)& (m)&  (m)&(m)\\
  \hline
  576&5.4&4.3&5.7&5.1\\
  1152&4.3&3.4&{\large\textcircled{\footnotesize 4.9}}&4.4\\
  2304&4.1&3.1&5.1&4.7\\
  \hline
\end{tabular}
\end{center}
%\vspace{-0.6cm}
\end{table}

In Fig. \ref{fig:RSRP}, we investigate the main hurdle in improving positioning accuracy based on RSRP values of the beams from the serving cell alone. The figures are plotted by choosing a UE randomly and then finding the UEs in its close proximity. We observe high correlation in serving cell beam IDs and their respective RSRP values for the UEs, which makes it harder to improve positioning accuracy of the UEs in close proximity. Fig. \ref{fig:nonoverlape} and Fig. \ref{fig:overlape} show 2 such examples where RSRPs for the strongest beam IDs are almost overlapping. After some further experimentation with various number of strongest beams from the serving cell with DNN (not reported in this paper), we observe a negligible difference in positioning accuracy by using 3 instead of 4 beams from the serving cell. As UE reporting is costly operation, we use RSRP values for at most 3 beams from the serving cell for further investigation in this sequel.

\begin{table}
\caption{Positioning Accuracy for with Fixed 3 beam RSRPs from the Serving Cell and Varying beam RSRPs from the Neighboring Cells}
%\vspace{-0.4cm}
\label{tab:serving and neighboring}
\begin{center}
\begin{tabular}{|c|c|c|c|c|c|}
\hline
  No. beams from &Training $\bar{E}$  &Training $\sigma$  &Test $\bar{E}$ &Test $\sigma$ \\
  neighbor cells& (m)& (m)&  (m)&(m)\\
  \hline
  0&4.3&3.6&4.8&4.7\\
  1&2.8&2.6&3.6&4.8\\
  2&2.2&1.5&{\large\textcircled{\footnotesize 3.0}}&3.4\\
  3&2.1&1.5&3.1&4.9\\
  \hline
\end{tabular}
\end{center}
%\vspace{-0.6cm}
\end{table}

Next, we propose to make use of RSRP values of the beams not only from the serving cell but from the neighboring cells as well. In Table \ref{tab:serving and neighboring}, we show position accuracy results by fixing the number of strongest beams from the serving cell to 3 and varying the number of beams from the neighbor cells. For example, using 2 strongest beams from 2 neighbor cells along with their beam and cell IDs give us 6 more features in addition to 7 features from 3 beam IDs, their respective RSRP values and cell ID of the serving cell. From Table \ref{tab:serving and neighboring}, we observe that mean accuracy of 3m is achieved by using data from 2 strongest beams from the neighboring cells in addition to 3 serving cell beams. This is considerable improvement over the case using data from beams only from the serving cell. This performance gain is achieved by adding more features and data for the DNN model training.

%%%%%%%%%%%%%%%%%%%%%%%%%%%%%%%%%%%%%%%%%55

\subsection{Cell-Specific DL Training}
\label{sect:Cell-specific}
Looking beyond feature engineering, we investigate the effect of change in location of training. In Section \ref{sec:network_level}, the DNN training was performed with all the data from several sites at one location server. In this section, we study positioning accuracy by training individual cell-specific DNN model for each cell. This results in removing the serving cell ID feature from the feature vector as it is known a-priori, called dimensionality reduction in ML literature. We preprocess data generated by the system level simulator to differentiate data for various cells. As a result, size of both training and test data may vary from cell-to-cell. For example, we have data for more than $10\times 10^3$ positions for cell number 14, while it is about $700$ for cell number 1. However, this variance in number od training examples does not affect results in our case as the resolution for Ray-tracing is kept fixed for all the cells regardless of their size.

%From Table \ref{tab:winprop}, we know that we had more than $122\times 10^3$ location samples for the cell level training gathered over all 24 sites. Lack of data could pose a major challenge in training of DNNs in Cell-Specific training.

\begin{table}
\caption{Cell-Specific-Training, Single Hidden Layer: Positioning Accuracy for data from both Serving and Neighboring Cells}
%\vspace{-0.4cm}
\label{tab:cell_specific 14}
\begin{center}
\begin{tabular}{|c|c|c|c|c|c|}
\hline
  No. beams from &Training $\bar{E}$  &Training $\sigma$  &Test $\bar{E}$ &Test $\sigma$ \\
  neighbor cells& (m)& (m)&  (m)&(m)\\
  \hline
  0&2.9&3.0&3.3&3.1\\
  1&2.3&2.0&2.8&5.1\\
  2&1.7&1.2&{\large\textcircled{\footnotesize 2.0}}&2.1\\
  3&1.4&1.1&1.9&1.7\\
  \hline
\end{tabular}
\end{center}
%\vspace{-0.6cm}
\end{table}

\begin{table}
\caption{Cell-Specific-Training, 2 Hidden Layers: Positioning Accuracy for data from both Serving and Neighboring Cells}
%\vspace{-0.4cm}
\label{tab:cell-specific_2layers_14}
\begin{center}
\begin{tabular}{|c|c|c|c|c|c|}
\hline
  No. &Training $\bar{E}$  &Training $\sigma$  &Test $\bar{E}$ &Test $\sigma$ \\
  Neurons& (m)& (m)&  (m)&(m)\\
  \hline
  32&2.2&1.6&2.4&1.6\\
  64&1.1&0.7&{\large\textcircled{\footnotesize 1.4}}&1.2\\
  128&1.2&0.8&1.5&1.4\\
  \hline
\end{tabular}
\end{center}
%\vspace{-0.5cm}
\end{table}

\begin{figure}
\centering
  	\includegraphics[width=3.5in]{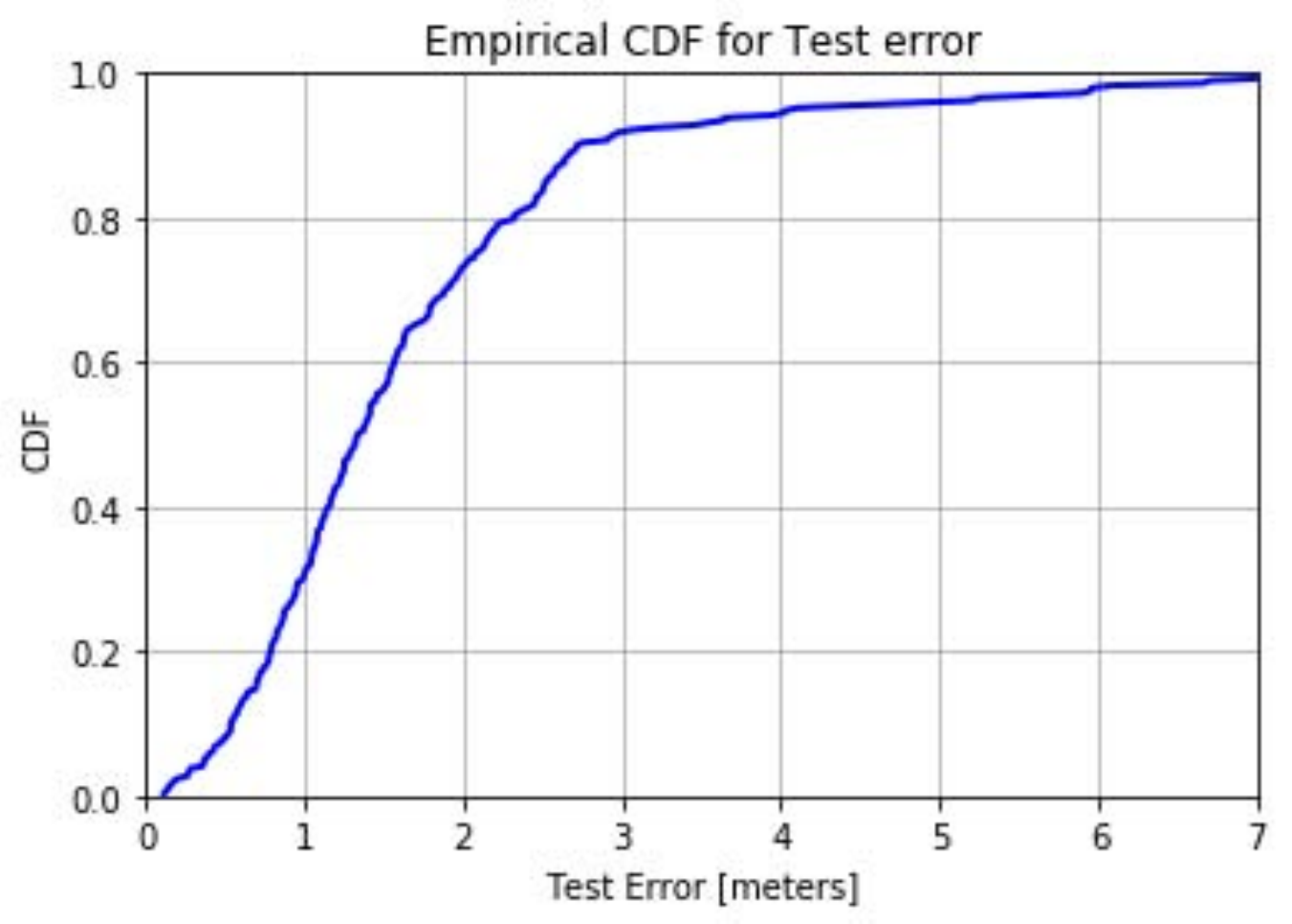}
  %\vspace{-0.4cm}
   \caption{CDF of positioning accuracy for the 2 hidden layer, cell specific DL training for cell 14.}
	\label{fig:error_CDF}
%\vspace{-0.4cm}
\end{figure}

In Table \ref{tab:cell_specific 14}, we study DNN with a single hidden layer and cell specific training for cell number 14. As previously observed, best feature vector uses data from 3 strongest beams from the serving cell. We further investigate accuracy by varying the number of beams from the neighboring cells for cell-specific training. Again, positioning accuracy improves by using data from more beams from the neighboring cells. In this case, the largest mean positioning accuracy is achieved for 3 neighboring cells; but difference is negligible as compared to 2 neighboring cell case. Looking at the cost of reporting information for this extra beam from the neighboring cell, we can safely assume that most of the performance gain is achieved with data from 3 serving and 2 neighboring cell beams. Remarkably, cell-specific training provides us mean positioning accuracy of 2m for 3 serving and 2 neighboring cell beams case which is considerably higher than the accuracy of 3m for the same configuration for network level training. This gain emerges from preprocessing step to get cell-specific data from the data set and removing\emph{ Cell ID} feature. To further achieve better results, we now look at further improving results by adding one more hidden layer to DNN. Table \ref{tab:cell-specific_2layers_14} shows that mean positioning error can be reduced to 1.4m by using 64 nodes in both hidden layers of DNN, a performance gain of 0.5m over the same parameters and single hidden layer case.
To elaborate further on the results, Fig. \ref{fig:error_CDF} shows cumulative distribution function (CDF) for this particular parameter set\footnote{CDF is the most commonly used criterion for reporting results in industry as it gives information about other percentile, e.g., $80\%$, $90\%$ percentile.}. The results illustrate that almost 30\% UEs achieve positioning error less than 1m and less than 10\% have error greater than 3m.

\begin{figure}
\centering
  	\includegraphics[width=3.5in]{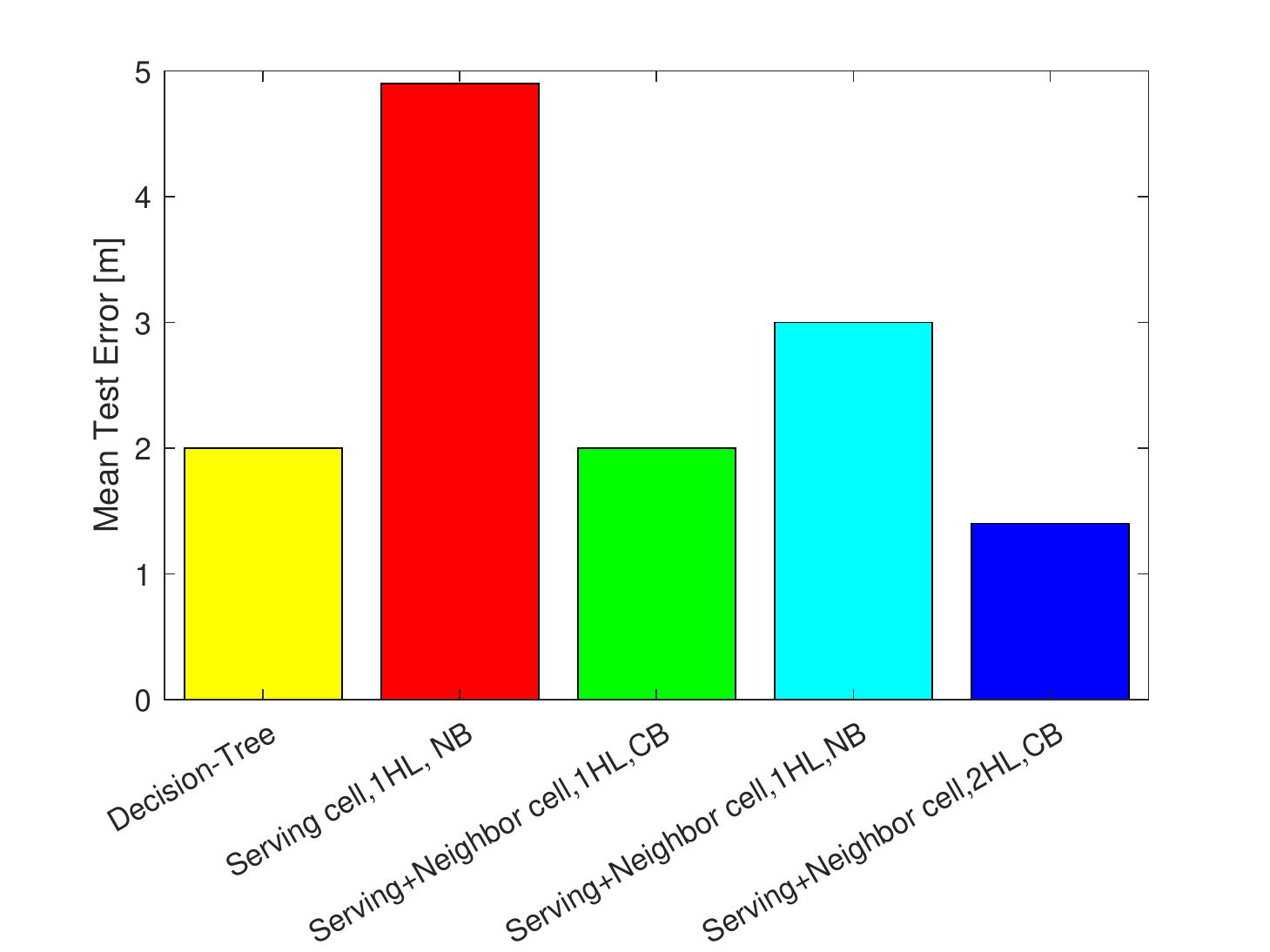}
   \caption{Performance comparison of Decision Tree Regressor and various DNN cases with different input features, number of hidden layers (HL) and architecture for training data, i.e. network based (NB) or cell-based (CB).}
	\label{fig:comp}
\end{figure}

%Fig \ref{fig:comp} shows positioning accuracy for combinations of serving cell and neighboring cell beams for Decision tree approach. The best results are again obtained for the case with 3 beams from the serving cell and 2 from the neighboring cells.
To conclude, we summarize all the results in Fig. \ref{fig:comp} and show that DNN with cell-based training using both serving as well as neighboring cell RSRP beam features provide the most accurate results. We include evaluation results for the baseline Decision-Tree Regression approach for \emph{network level} training. Network level training eliminates the need for cell-level training servers without much loss of performance as decision tree regressor can easily differentiate cells based on their IDs at first level of branching as depicted in Fig.~\ref{fig:tree}. The input features for the baseline case are the same as for DNN approach, i.e., 3 beams from the serving cell and 2 from the neighboring cells. We get mean positioning error of 2.1m which is less than the mean positioning error obtained by DNN approach.

\section{Architecture Support for ML-Assisted Use Cases}
\label{sec:arch_5G}
Based on our performance analysis in previous section, we discuss architecture level aspects to implement ML solutions in 5G networks. In relation to measurement data collection and network support, it is important to answer a few fundamental questions for any use case enabled with ML assistance:
\begin{itemize}
\item \textbf{Data sources}: How and from where to collect sufficient data for ML training and inference?
\item \textbf{Training and inference hosts}: What are the 5G system (5GS) entities to place the ML training and inference modules?
\item \textbf{Required enhancements}: Does the current 5GS architecture support the required data collection? Which 5GS architecture enhancements would be needed?
\end{itemize}
To answer the above for the UE positioning use case as an example, we have primarily looked at the fundamental architecture problems associated with implementation of ML and specifically supervised learning techniques in 5G radio access network (RAN).

\begin{figure*}
\centering
  	\includegraphics[width=5.0in, scale= 1]{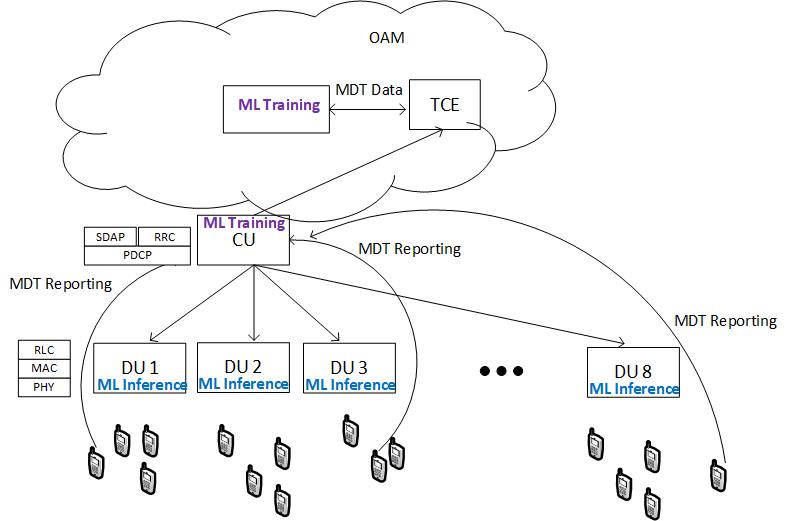}
   \caption{Proposed ML-Assisted RRM architecture. As illustrated on the left side of the figure (gNB$^1$ (CU$^1$)) ML Training can be performed in gNB-CU, and OAM, while ML Inference can be performed in the gNB-DU or gNB-CU. The exact possibilities on ML Training and Inference for the specific use cases of Positioning are given at the right side of the figure (gNB$^2$ (CU$^2$)).}
	\label{fig:arch}
\end{figure*}

5G NR architecture introduced a split RAN architecture for a 5G NodeB (gNB), where a gNB consists of a Centralized Unit (gNB-CU) and one or more Decentralized Units (gNB-DU) \cite{TS38.401}. In the split architecture adopted by 3GPP, the lower layers of the protocol stack reside at the gNB-DU and the higher layers, including the radio resource control (RRC), are part of the gNB-CU  \cite{TS38.401}. An example of split architecture with 2 gNB-CUs hosting 3 gNB-DUs each, is illustrated in Figure \ref{fig:arch}. Given this split architecture, all the RRC specific messages and configurations are performed at the gNB-CU.
To be able to monitor and maintain network operation, network configures UEs with measurement reports. Measurement reports can be configured from RRC in which case the configured UE measurements are sent to the gNB-CU. One type of UE reports is obtained by the network through configuration of MDT reports. Additionally, network can configure radio measurements, involving the physical and MAC layers, as well as their reporting. Those measurements are either periodic in nature or sent upon trigger and are available at the gNB-DU hosting the lower layers in the form of channel quality indicator (CQI) reports.

\begin{table*}
\begin{center}
%\footnotesize

\caption{Summary of ML-assisted UE Positioning Use Case}

%\vspace{-0.3cm}
\begin{tabular}{ll}
\toprule
 \textbf{Requirements for ML}& \textbf{5G Network Options}\\
\midrule
\textbf{Data Source}& \textbf{Training:} UE measurements (RSRP, location); other options (AoA, ToA)\\
   &  Inference: UE measurements (RSRP); other options (AoA, ToA)\\
   & \\
   \textbf{Training Host}& \textbf{option 1:} gNB-CU \\
  &\textbf{option 2:} OAM     \\
&\\
  \textbf{Inference Host}&gNB-DU  \\
  &  \\
\textbf{Required Enhancements}& {\textbf{UE:} periodic Immediate MDT reports with low overhead}\\
  & {\textbf{OAM:}  ML-data processing and storage from large number of cells and UEs}\\
  & {\textbf{gNB/gNB-CU:}  ML-data processing and storage, transfer to/ from other gNBs}\\
  & {\textbf{OAM and gNB/gNB-CU communication:} ML-data transfer between gNB and OAM/TCE}\\
  &\\
\bottomrule
\end{tabular}

\label{tab:summary}

\end{center}
%\vspace{-0.4cm}
\end{table*}

%\begin{table*}
%\caption{Summary of ML-assisted UE Positioning Use Case}
%\label{tab:summary}
%\begin{center}
%\begin{tabular}{|l||l|}
%\hline
%
%  \hline
%  \textbf{Requirements for ML}& \textbf{5G Network Options}\\
%   & \\
%  \hline
%  \hline
%  \emph{Data Source}& \textbf{Training:} UE measurements (RSRP, location); other options (AoA, ToA)\\
%   &  \emph{Inference:} UE measurements (RSRP); other options (AoA, ToA)\\
%   & \\
%  \hline
%  \emph{Training Host}& \textbf{option 1:} gNB-CU \\
%  &\textbf{option 2:} OAM     \\
%  &   \\
%
%  \hline
%  \emph{Inference Host}&gNB-DU  \\
%  &  \\
%
%  &\\
%  \hline
%
%  \emph{Required Enhancements}& {\textbf{UE:} periodic Immediate MDT reports with low overhead}\\
%  & {\textbf{OAM:}  ML-data processing and storage from large number of cells and UEs}\\
%  & {\textbf{gNB/gNB-CU:}  ML-data processing and storage, transfer to/ from other gNBs}\\
%  & {\textbf{OAM and gNB/gNB-CU communication:} ML-data transfer between gNB and OAM/TCE}\\
%  &\\
%  \hline
%\end{tabular}
%\end{center}
%\end{table*}

\subsection{Measurement Collection for Running ML Solutions}
Supervised ML solutions comprise two phases, the training phase where measurements are used to train an ML model and the inference phase where the model is executed. In order to train an ML model, it is important to obtain "sufficiently" large number of high-quality measurements for training an ML model.
In Long Term Evolution cellular systems (LTE), 3GPP introduced Minimization of Drive Tests (MDT) \cite{TS37.320} as a methodology to support network planning and optimization by detecting and troubleshooting erroneous network situations, such as coverage holes and weak coverage in the uplink or downlink directions to name a few.\footnote{MDT is a feature introduced in 3GPP Rel. 10 which allows operators to collect data from UEs, e.g., radio measurements and location information, in order to improve network performance.} To achieve that, MDT requests, manages and processes a (potentially) large amount of data.
Even though MDT was originally introduced in LTE, it has currently been under NR standardization.
Since MDT requests measurements from the UEs, it has been considered a viable solution to obtain radio measurements for ML model training. Similarly, for the ML inference phase, (near) real time data needs to be collected from the MDT and an appropriate action is performed as a result of ML inference.

In MDT, the network can configure a UE with specific measurements. There are two modes for MDT measurements: Logged MDT and Immediate MDT. In Logged MDT, the network configures a UE in RRC Connected State with a measurement configuration. The UE logs measurements according to its configuration while in RRC \textit{Idle} or RRC \textit{Inactive} states. Then, UE indicates data availability to the network when it returns to RRC \textit{Connected} state so that the network can retrieve it. In LTE, logging was only periodical but in NR event-based logging is also introduced to address "out of coverage" detection.

Since MDT handles UE measurements, it has been considered as a viable methodology for training of ML models. Also, both Immediate and Logged MDT can be useful in providing UE measurements for UE localization since each method targets UEs in a different state; in Immediate MDT measurements are provided to the network by UEs in RRC Connected state while in Logged MDT a UE logs measurements while in RRC Idle or Inactive states. By using both methods to obtain measurements a larger number of UEs can be targeted providing a larger and more diversified set of measurements in short time.

Despite its natural connection to data collection for ML, MDT is an optional feature and does not bind UE manufacturers with implementing it. Furthermore, user consent issues may arise where a user may not be willing to provide data to the network exposing for instance its location or other user-specific information. Thus, it is still challenging to obtain the required measurements from UEs. Besides, the number of measurements required to obtain meaningful training results can be large and depends on the performance sought by the ML algorithm. For example, in UE positioning the number of measurements collected for basic best effort positioning can be much less than those when some guarantee on the positioning accuracy is expected. Furthermore, since obtaining and reporting of UE measurements consumes UE resources (e.g., in terms of battery power, memory, etc.), there should be a balance in the number of measurements that the UE is configured to report to the network and in the performance gains.

To address these problems above, network can further use measurements internally collected and accumulated. For instance, downlink CQI feedback can be collected by the gNB-DU and could be used in combination with RRC-based reporting methods to help network with the data collection. Using such methods would relieve some of the requirements on UE reporting for data collection. On the other hand, this will come at the cost of a higher complexity on collecting all the measurements on a single network entity since some of the measurements may be available at the gNB-CU and others at a gNB-DU.

All above factors render the measurement collection process to be complex. In this paper, we assume that UEs can provide the requested measurements to the network, at least via the standardised RRC measurement reports, and we discuss various options for the ML training and inference functions.

\subsection{ML Training and ML Inference Modules}
As we discussed in the previous section, the training of an ML model requires a potentially large number of measurements. Thus, it is natural to assume that the training of an ML model will be placed at a network entity where the measurements are available. In the 5G architecture, MDT measurements can be reported by the UEs to the gNB-CU. This implies, that gNB-CU is a network entity that may receive combined UE measurements from all UEs under this gNB-CU. A gNB may log these measurements and forward the logs to the Trace Collection Entity (TCE) that resides in the Operations and Management (OAM) of the network. Thus, 5G architecture provides two natural locations to perform the ML training, the gNB-CU and the OAM. Performing ML training at the gNB-CU provides centralized optimization over the received measurements collected by the UEs connected to it. On the other hand, ML training at the OAM could additionally provide even more global optimization by combining measurements obtained across gNBs.

%Another possible location for ML training is the UE. The UE performs high rate radio measurements with relatively high accuracy, thus it can use these to train an ML model if it is downloaded to, or implemented in the UE. More advanced methods also exist that use a combination of ML training both at the UE and at the gNB.

For ML based positioning use case, non-real time, or near real-time, inference is sufficient for most of the applications, which can be run in the gNB (DU or CU). Where to run ML inference exactly, depends on the ML-assistance use case. Specifically, when it comes to ML for localization, lower layer UE measurements are used for the inference. It is therefore natural that the inference is placed at the gNB-DU. Figure \ref{fig:arch} depicts the possible locations where ML training and ML inference can be placed within a split 5G-RAN architecture.

%%%%%%%%%%%%%%%%%%%%%%%%%%%%%%%%%%%%%%%%%%%%%%%%%%%%%

\section{Summary: Design Considerations}
\label{sec:arch_support}

In Table \ref{tab:summary}, we summarise our findings, and the proposed answers to the main questions raised in Section \ref{sec:arch_5G}.

The UE positioning relies on the UE downlink radio measurements and their corresponding RRC reports or MDT reports to the serving gNB (or gNB-CU). ML-assisted RRM algorithms can accommodate multiple types of input data sources simultaneously, which can further enhance their inference capabilities. Examples for such inputs are: RSRP measurements, UE location information, etc, while others like cell load measurements, UE traffic type can as well be used for other ML based use cases.

These types of ML-assisted mechanisms must rely on a minimum 'quality' of the input data in order to ensure good performance. This is not a trivial task, especially when considering inputs generated in systems not fully controlled by the 3GPP network and its RRM algorithms.

Training of the proposed ML-assistance algorithms for the studied UE positioning use case can be done at the gNB and/ or at the UE. More advanced training can combine joint training at the gNB and at the UE. Furthermore, since MDT measurements are forwarded from the gNB to the OAM, ML training at the OAM can further take advantage of the higher computational power it offers. Additionally, since OAM is a reception point of MDT reports from multiple gNBs, it can further provide a centralized system-wide ML solution. ML training at OAM is also a viable solution for enhanced localization purposes.
Particularly for the localization use case, the ML inference is performed at the gNB-DU, since this is a network entity with knowledge on lower layer information, e.g., beam information and RSRP values.

%In any of the configuration above, new radio signalling elements between UE and gNB, including UE ML capability selection and UE ML configuration are needed. Downloading specific ML models to the UE, which have been designed in the network (e.g. from OAM) is one possible solution.

The ML-assistance, data processing and data transfer functionalities need to be available in the gNB (DU and/or CU) or OAM, either directly implemented in these entities, or in separate host entities with direct interfaces to the gNB and OAM \cite{ITU_recommend3172}.

\section{Conclusions}
\label{sec:conclusions}
In this article, we have comprehensively analyzed UE positioning use case for application of machine learning for performance enhancement in 5G networks. Performance of UE localization use case is evaluated for various input features, network configuration for model training and neural network architecture; and UE positioning accuracy of 1-1.5m is observed for LOS UE measurements. Furthermore, the challenges in implementing machine learning solution for the studied use case are discussed by 5G-RAN architecture point of view. Various options for ML training and inference hosting are considered and their trade-off in terms of complexity is presented. We outline our vision for architectural level support in 5G (leading to 6G) networks for integrating ML for the promising use cases where ML solutions can achieve significant gains in future ML-based network architectures.
\section{Acknowledgements}
The authors would like to thank Daejung Yoon and Afef Feki for having useful discussions on UE positioning in 3GPP that helped to improve content of this paper.

\bibliographystyle{IEEEtran}
\bibliography{bibliography}

\end{document}